\documentclass[letter]{aa}
\usepackage{graphicx}
\usepackage{natbib}
\usepackage{txfonts}
\begin{document}
   \title{Penumbral models in the light of Hinode spectropolarimetric
   observations}

   \author{J. Jur\v{c}\'{a}k
          \inst{1,2}
          \and
          L.R. Bellot Rubio\inst{1,3}}

   \institute{National Astronomical Observatory of Japan, 2-21-1 Osawa,
Mitaka, Tokyo 181-8588, Japan
        \and
        Astronomical Institute of the Academy of Sciences, Fricova
298, 25165 Ond\v{r}ejov, Czech Republic
         \and
          Instituto de Astrof\'{\i}sica
de Andaluc\'{\i}a (CSIC), Apdo.\ Correos 3004, 18080 Granada, Spain
             }

   \date{Received September 15, 1996; accepted March 16, 1997}


  \abstract
   {}
   {The realism of current models of the penumbra is assessed by comparing their predictions
    with the plasma properties of penumbral filaments as retrieved from spectropolarimetric
    observations.}
   {The spectropolarimeter onboard Hinode allows us to distinguish for the first time the fine
   structure of the penumbra. Therefore, we can use one-component inversions to obtain the
   stratifications of plasma parameters in each pixel. The correlations between the plasma
   parameters and the continuum intensity are studied.}
   {We find that, in the outer penumbra, the stronger flows and higher values of magnetic field
   inclination tend to be located in dark filaments.
   This finding does not seem to be compatible with the scenario of a field-free gappy penumbra.}
   {}

   \keywords{ Sun: sunspots --
              Sun: photosphere  --
              Techniques: polarimetric
               }

   \maketitle
%

\section{Introduction}

There are two models of the penumbral fine structure under discussion nowadays.
The first is the uncombed model proposed by \citet{Solanki:1993} and further
developed by \citet{vmp:2000} to explain the broad-band circular polarisation
observed in sunspots.  It envisages the penumbra as a collection of horizontal
flux tubes embedded in a more vertical and stronger background field.
\citet{Schlichenmaier:1998} made simulations of the temporal evolution of such
flux tubes. Initially placed at the sunspot magnetopause, they rise by magnetic
buoyancy and quickly become horizontal in the photosphere. At the same time, a
strong Evershed flow develops due to the gas pressure gradient that builds up
during the rise. The tubes are hotter than the surrounding plasma and appear as
bright penumbral filaments which gradually cool down due to radiative losses,
i.e. they become dark further from the bright penumbral grains where the tubes
cross the $\tau = 1$ layer. Hereafter, this model will be called the rising
flux tube model (RFT).

The second model was suggested by \citet{Spruit:2006}. These authors argued
that the RFT model cannot produce sufficient heating of the penumbra. As an
alternative, they proposed a field-free gap model (FFG) which consists of
field-free material protruding into the penumbral magnetic field from below.
With such a configuration, the bright filaments would be heated all along their
lengths and not only at the point of emergence as in the case of the RFT model.
While the FFG model does not offer any explanation for the Evershed flow, there
is only one place where the flow can reside, namely on top of the field-free
gaps, since this is the sole region in the model where horizontal fields exist.

\begin{figure}[!b]
  \centering
  \includegraphics[width=\linewidth]{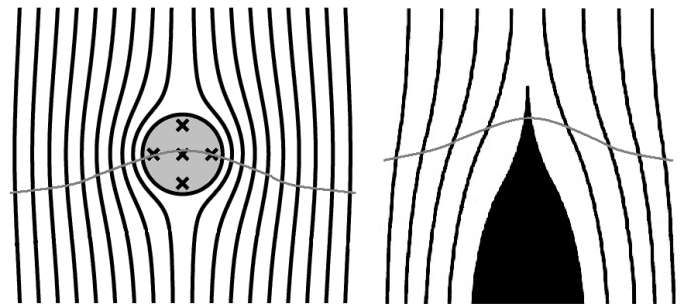}
  \caption{Sketch of a RFT (left) and a FFG (right) surrounded by the
  background magnetic field of the penumbra. The gray line
  represents the possible position of the optical depth unity layer.}
  \label{models}
\end{figure}

Both models have problems from a theoretical point of view which need to be
clarified \citep[see][and references therein]{Bellot:2007}. However, as
illustrated in Fig.~\ref{models}, RFTs and FFGs should produce similar
stratifications of the plasma parameters if they are placed near $\tau = 1$
(marked by the gray line). The two models predict weaker and more inclined
fields low in the atmosphere, along with increased line-of-sight (LOS)
velocities. Using Hinode spectropolarimetric data, \citet{Jurcak:2007}
found this configuration in the bright filaments of the inner penumbra.

\begin{figure}[!t]
 \centering
 \includegraphics[width=\linewidth]{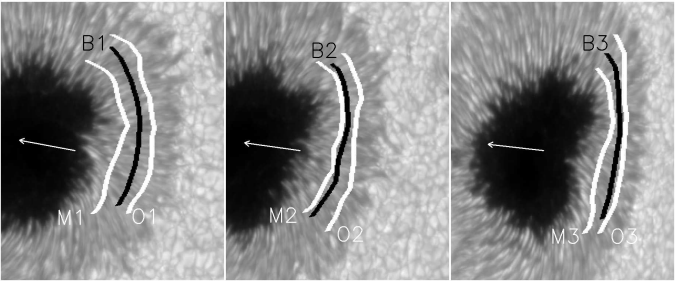}
\caption{Maps of continuum intensity at 630.3 nm reconstructed from the SP
raster scans. North is up and West to the right. The maps cover the limb side
part of the penumbra in evolving AR 10930. The arrows point to the disc centre.
The black and white bands indicate the areas selected for analysis.}
 \label{imaps}
\end{figure}

In this Letter we concentrate on the mid and outer penumbra, where
differences between the models could be more pronounced. According to
the FFG model, horizontal fields and increased LOS velocities should
be associated with bright areas everywhere in the penumbra. By
contrast, the RFT model expects such characteristics in the dark
filaments of the outer penumbra.

\section{Observations and data analysis}

The data analysed here were obtained using the spectropolarimeter
\citep[SP;][]{Tarbell:2007} of the Solar Optical Telescope \citep{Tsuneta:2007}
onboard the Hinode satellite \citep{Kosugi:2007}. This instrument measures the
Stokes profiles of the two iron lines at 630.15~nm and 630.25~nm.

Normal SP scans of AR 10930, providing a spatial resolution of
0\farcs32 and a noise level of 10$^{-3} I_{\rm c}$, were taken on
December 13, 14, and 15, 2006, when the spot was located at
heliocentric angles of 30$^\circ$, 41$^\circ$, and 54$^\circ$,
respectively. The three panels of Fig.~\ref{imaps} show the limb-side
penumbra of AR 10930 as reconstructed from the continuum intensities
observed redward of the 630.25~nm line.  The Stokes $V$ profiles
emerging from the penumbra exhibit significant area asymmetries due to
the existence of gradients of the plasma parameters in the
line-forming region. We can clearly see the bright and dark filaments
in the continuum maps. Their signals are not mixed as in the case of
ground-based observations, implying that the penumbral fine structure
is distinguished in spectropolarimetric measurements for the first
time. Higher spatial resolution observations of the penumbra exist
(e.g., Scharmer et al.\ 2002; Rouppe van der Voort et al.\ 2004;
Rimmele 2004; Langhans et al.\ 2005; Bellot Rubio et al.\ 2005;
S\'anchez Almeida et al.\ 2007), but they correspond to filtergrams,
Dopplergrams, magnetograms, and spectroscopic measurements from which
the magnetic field vector cannot be determined.

Each day we selected three 5-pixel-wide azimuthal cuts. As illustrated in
Fig.~\ref{imaps}, they sample the mid and outer parts of the penumbra.  The
bands were chosen so as to minimise any trend of the continuum intensity along
them, except for small-scale fluctuations caused by the alternating bright and
dark filaments. In total, they contain 5862 pixels.

The data reduction is described in detail in \citet{Jurcak:2007}. The zero
point of the velocity scale is taken to be the line core position of the mean
quiet Sun intensity profile, computed separately for each slit. The observed
Stokes spectra have been inverted using the SIRGAUS code \citep{Bellot:2003},
which is a modified version of SIR \citep{Cobo:1992}. This code presumes the
existence of a Gaussian perturbation (GP) in the stratifications of plasma
parameters somewhere in the line-forming region. Given the high angular
resolution of the Hinode measurements, we only consider one-component model
atmospheres.

The inversion code looks for the best solution in a space of 13 free
parameters. Six of them define the physical conditions of the
unperturbed atmosphere: we use two for the temperature ($T$), and one
for the field strength ($B$), inclination ($\gamma_{\rm LOS}$),
azimuth ($\phi_{\rm LOS}$), and LOS velocity ($v_{\rm LOS}$). The
width and position of the GP is the same for all plasma parameters,
which adds two more free parameters. The rest of parameters are the
amplitudes of the perturbation in the different physical
quantities. As initial values we adopt $\Delta T = +800$ K, $\Delta B
= -500$ G, $\Delta \gamma_{\rm LOS} = -30^\circ$, $\Delta \phi_{\rm
LOS} = -5^\circ$, and $\Delta v_{\rm LOS} = +3$~km~s$^{-1}$. These
initial parameters imply weaker and more inclined fields associated
with stronger Evershed flows, so they can model both RFTs and
FFGs. The magnetic field inclination and azimuth retrieved by the
inversion code are line-of-sight values. We transformed these
variables to the local reference frame. In what follows, the field
inclination ($\gamma$) represents the angle between the magnetic field
vector and the inward normal to the solar surface.

\section{Choice of plasma parameters}

As explained by \citet{Jurcak:2007}, we perform four inversions per pixel with
different starting heights of the GP, and select the one that shows the
smallest value of the merit function (sum of squared differences between
observed and synthetic profiles). It sometimes happens that the Stokes spectra
emerging from a given pixel can be fitted equally well by GPs located either in
deep layers or in the mid photosphere. In these cases it is difficult to decide
which solution is closer to reality on the basis of the merit functions alone,
as they are usually of comparable value. Although the resulting stratifications
are different, they share some similarities: variations in the position of the
GP mainly change the amplitude and sign of the magnetic field strength
perturbation, but the sign and amplitude of the GP for the LOS velocity and the
field inclination are almost independent of the height of the Gaussian.

We take advantage of this fact by adopting the part of the solution which is
similar in all cases, i.e., the average values of the plasma parameters in the
line-forming region. Specifically, we compute for each pixel the mean value of
the LOS velocity, field inclination, and field strength in the range from $\log
\tau = -0.5$ to $\log \tau = -2.5$. This is the region where the Stokes
profiles of the \ion{Fe}{i} 630.2 nm lines are more sensitive to the physical
conditions of the atmosphere (Cabrera Solana et al.\ 2005).

From the obtained values we subtract the locally averaged plasma
parameters, as defined by a running mean over 100 pixels. The
resulting fluctuations of continuum intensity, LOS velocity, field
inclination, and field strength are used to test the predictions of
the RFT and FFG models.

\section{Results}

\begin{figure}
 \centering
 \includegraphics[width=0.99\linewidth]{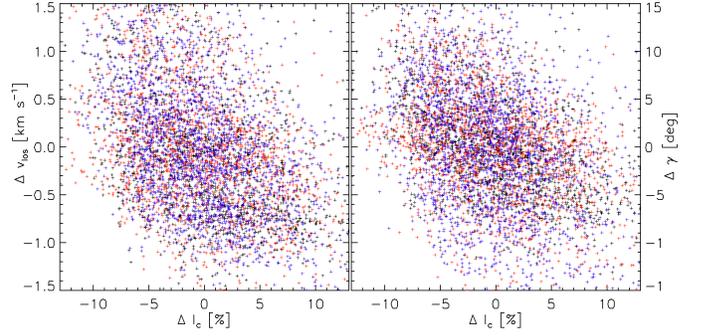}
\caption{Scatter plots of LOS velocity ({\em left}) and inclination ({\em
right}) fluctuations vs. continuum intensity fluctuations. The black, red, and
blue symbols represent points from the M, B, and O bands, respectively.}
 \label{scatter_plot}
\end{figure}

\begin{figure*}
 \centering
 \includegraphics[width=16.5cm]{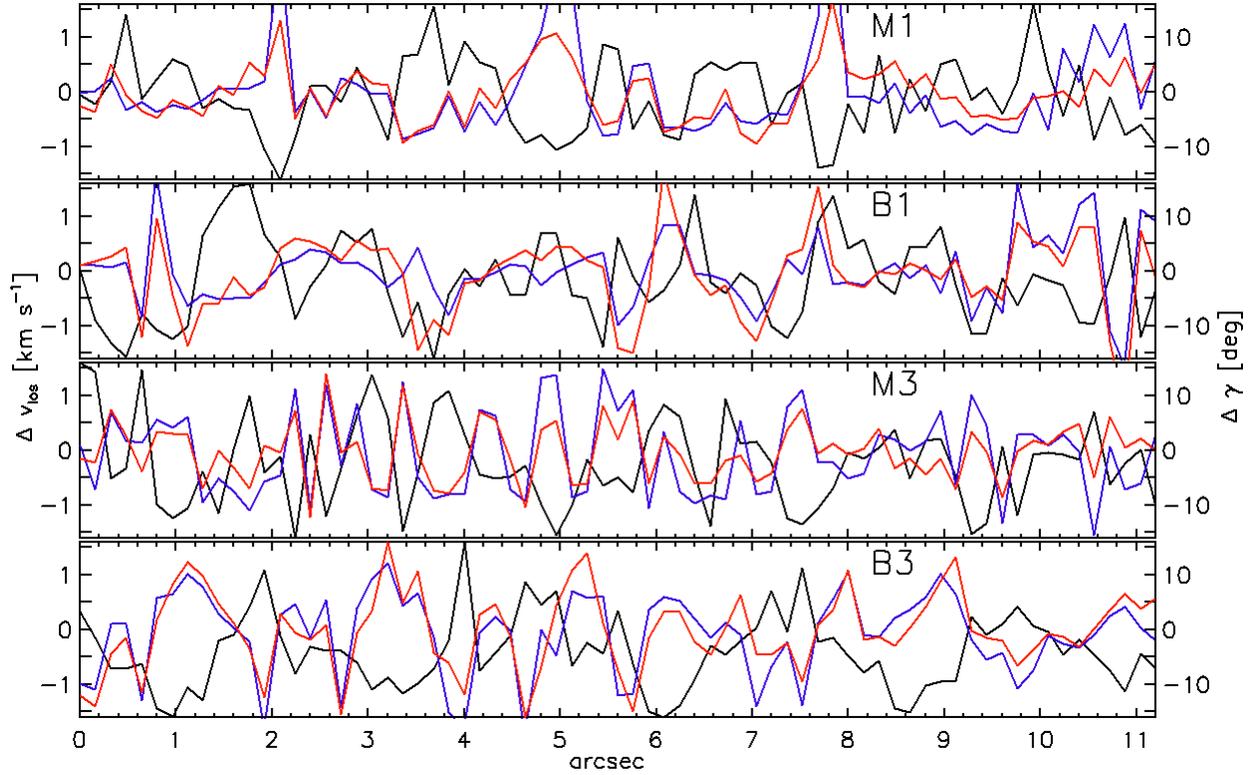}
\caption{Continuum intensity (black), LOS velocity (blue), and magnetic field
inclination (red) fluctuations along the selected paths. The plots show, from
top to bottom, parts of the M1, B1, M3, and B3 bands.} \label{corr}
\end{figure*}

Figure~\ref{scatter_plot} shows scatter plots of LOS~velocity and
magnetic field inclination fluctuations vs. continuum intensity
fluctuations. There is a clear tendency for weaker flows and more
vertical fields in brighter penumbral areas. However, this type of
plot does not reflect how the plasma parameters are associated with
penumbral filaments. Therefore, in Fig.~\ref{corr} we give examples of
continuum intensity fluctuations (black lines) and the corresponding
changes of LOS~velocity (blue lines) and inclination (red lines) along
one-pixel-wide azimuthal cuts extracted from the selected bands.

Table~\ref{tab} summarises the correlation coefficients obtained in
the nine bands, with errors representing 95\% confidence
intervals. The values shown in the second and third columns confirm
the visual impression from Figs.~\ref{scatter_plot} and~\ref{corr}
that, in the mid and outer penumbra, the stronger flows and the more
inclined fields usually occur in dark filaments. The average
correlation coefficient between $\Delta v_{\rm LOS}$ and $\Delta
I_{\rm c}$ amounts to $-0.40$, which is quite high compared to
previous analyses based on spectropolarimetric data
\citep[e.g.,][]{cwp:2001, Solanki:2003}. Enhanced LOS velocities
in dark filaments were also found from high-resolution filtergrams
with comparable or even better correlations
\citep[e.g.,][]{Schlichenmaier:2005, Ichimoto:2007}.

The third column of Table~\ref{tab} shows the correlation coefficients
between $\Delta \gamma$ and $\Delta I_{\rm c}$. In this case the
average correlation is $-0.40$ ($-0.46$ if we do not take into
account bands B1 and O1, which are significantly worse). Thus, there
is a tendency for the more horizontal fields to be associated with
dark penumbral areas. Previous analyses reported either no significant
correlation between these parameters in the outer penumbra
\citep{Title:1993, Stanchfield:1997}, or larger field inclinations in
dark filaments \citep[e.g.,][]{Rimmele:1995, cwp:2001, Langhans:2005}.
Usually, the correlation coefficients quoted in those works are
smaller (in absolute value) than 0.2. Only when the correlation is
analysed at small spatial scales do the coefficients become similar to
(or higher than) the ones reported here.

\begin{table}
\centering
\caption{Correlation coefficients between fluctuations of LOS velocity
and continuum intensity (second column), field inclination and
continuum intensity (third column), field inclination and LOS velocity
(fourth column), and field inclination and strength (fifth column),
along the nine bands marked in Fig.~2. Errors represent 95\%
confidence intervals.}

\label{tab}
\begin{tabular}{@{\hspace{1mm}}c@{\hspace{3mm}}c@{\hspace{3mm}}c@{\hspace{3mm}}c@{\hspace{3mm}}c@{\hspace{1mm}}}
\hline\hline\noalign{\vskip1mm}
  & $\Delta v_{\rm LOS}$ & $\Delta \gamma$  &  $\Delta \gamma$ & $\Delta \gamma$ \\
Cut & vs. $\Delta I_{\rm c}$ & vs. $\Delta I_{\rm c}$ & vs. $\Delta v_{\rm LOS}$ & vs. $\Delta B$\\
\noalign{\vskip1mm} \hline\noalign{\vskip1mm}
M1 &    $-0.56\pm0.06 $&  $-0.55\pm0.06 $&    $0.82\pm0.03$&    $-0.78\pm0.04$\\
M2 &    $-0.23\pm0.08 $&  $-0.44\pm0.07 $&    $0.78\pm0.03$&    $-0.75\pm0.04$\\
M3 &    $-0.44\pm0.07 $&  $-0.40\pm0.07 $&    $0.78\pm0.03$&    $-0.69\pm0.04$\\
B1 &    $-0.34\pm0.08 $&  $-0.15\pm0.09 $&    $0.81\pm0.03$&    $-0.72\pm0.04$\\
B2 &    $-0.36\pm0.07 $&  $-0.53\pm0.05 $&    $0.67\pm0.04$&    $-0.71\pm0.04$\\
B3 &    $-0.41\pm0.07 $&  $-0.43\pm0.06 $&    $0.70\pm0.04$&    $-0.46\pm0.06$\\
O1 &    $-0.30\pm0.07 $&  $-0.18\pm0.07 $&    $0.86\pm0.02$&    $-0.30\pm0.07$\\
O2 &    $-0.42\pm0.06 $&  $-0.44\pm0.06 $&    $0.77\pm0.04$&    $-0.33\pm0.07$\\
O3 &    $-0.55\pm0.05 $&  $-0.45\pm0.06 $&    $0.60\pm0.04$&    $-0.30\pm0.06$\\

\hline
\end{tabular}
\end{table}

The two last columns of Table~\ref{tab} quantify the relation between
the changes of magnetic field inclination and the LOS velocity and
field strength.  It is generally accepted that the Evershed flow
resides in areas of horizontal fields. With correlation coefficients
as high as 0.7--0.8, the fourth column of Table~\ref{tab} just confirms
this fact. The fifth shows that the more inclined fields are weaker,
although the correlation coefficients decrease significantly from the
mid to the outer penumbra. This implies a smaller difference in
magnetic field strength between areas with horizontal fields (or flow
channels) and the surrounding atmosphere in the outer penumbra, which
is not surprising as both the RFT and FFG models predict an
equalisation of the field strength with increasing distance from the
umbra. A smaller field strength difference between inclined flow
channels and surrounding plasma has also been inferred from
two-component Stokes inversions \citep[e.g.,][]{Bellot:2004,
Borrero:2006} and analyses of the net circular polarisation observed
in the penumbra at high angular resolution \citep{tritchler:2007}.

The plasma parameter averages made over the line-forming region do
not tell us much about the actual properties of the flow channels
represented by the GP.  Therefore, we also studied the general
behaviour of the peak values of the GPs of LOS~velocity and magnetic
field inclination. We found that the flow becomes faster with
increasing distance from the umbra (on average it is 2~km~s$^{-1}$
faster in the O cuts than in the M cuts).  This increase confirms
previous reports of enhanced flow velocity with increasing distance
from the umbra \citep[e.g.,][]{Rimmele:1995, cwp:2001,Solanki:2003,
Bellot:2004, Borrero:2006}, and is also predicted by simulations of
moving penumbral tubes \citep{Schlichenmaier:1998}.

The maximum values of the LOS velocity are around 10~km~s$^{-1}$ near
the symmetry line in the O2, B3, and O3 cuts. Even if such velocities
are overestimated because they represent the peak amplitudes of the
GP, the absolute flow velocity must be very high due to projection
effects. These velocities are compatible with the supersonic flows
predicted by the simulations of \citet{Schlichenmaier:1998} in the
outer penumbra.

The average magnetic field inclinations in the flow channels are 85$^\circ$,
100$^\circ$, and 105$^\circ$ in the M, B, and O cuts, respectively. This means
that the flow channels are inclined upward in the mid penumbra and point down
in the outer penumbra (the polarity of the spot is negative). Magnetic fields
returning to the solar surface in the outer penumbra were already reported
by, e.g., \citet{cwp:2001} and \citet{Bellot:2004}.

\section{Discussion and conclusions}

The most important result of our analysis is that we do not find
positive correlation coefficients between $I_{\rm c}$ and $\gamma$ or
$v_{\rm LOS}$. This indicates that bright penumbral filaments show the
more vertical fields and weaker flows, in contradiction with the
predictions of the FFG model. The horizontal fields and stronger flows
tend to be associated with dark filaments, as has been reported
previously from lower resolution data. The fact that the correlation
coefficients between $I_{\rm c}$ and $\gamma$ or between $I_{\rm c}$
and $v_{\rm LOS}$ are not $-1$ can be easily explained by the mixing
of flux tubes and background magnetic fields. In the mid and outer
penumbra, dark areas may represent either the darkened tails of RFTs
or background fields.  The latter are characterised by more vertical
fields and weaker flows. Such an opposite behaviour decreases the
correlation coefficients, which nevertheless remain negative because
of the overall dominance of dark RFTs in the outer parts of the
penumbra.  This argument does not influence the correlation
coefficients between $\gamma$ and $v_{\rm LOS}$, which are indeed much
higher than those involving $I_{\rm c}$.

\citet{Schlichenmaier:2005} and \citet{Ichimoto:2007} came to a
similar conclusion (regarding the velocities) from simpler
interpretations of filtergrams and spectropolarimetric data,
respectively. They also reported that the horizontal fields and the
flows are associated with bright filaments in the innermost penumbra
\citep[see also][]{Hirzberger:2005,Jurcak:2007}, where the dark
areas most likely represent background magnetic fields. Thus,
the correlation coefficients between continuum intensity and
LOS velocity/inclination change sign from the inner to the
outer penumbra, due to the fact that RFTs predominantly show up as bright
structures in the inner penumbra but darken toward the outer edge of
the spot. This sign reversal cannot be explained by the FFG model. The
existence of magnetic field lines pointing to the solar interior in
the outer penumbra, also confirmed by our observations, represents
another problem for the FFG model \citep{Bellot:2007}.

The RFT simulations of \citet{Schlichenmaier:1998} seem to be able to
explain all the results of our analysis, including the sign reversal
of the correlation coefficients and the larger mean value of the LOS
velocity in the outer penumbra as compared with the mid
penumbra. Overall, this suggests that the RFT model provides a better
description of the penumbra than the FFG model.

\begin{acknowledgements} This work has been enabled thanks to the
funding provided by the Japan Society for the Promotion of Science. Hinode is a
Japanese mission developed and launched by ISAS/JAXA, with NAOJ as domestic
partner and NASA and STFC (UK) as international partners. It is operated by
these agencies in cooperation with ESA and NSC (Norway). The computations were
carried out at the NAOJ Hinode Science Center, which is supported by the
Grant-in-Aid for Creative Scientic Research The Basic Study of Space Weather
Prediction from MEXT, Japan (Head Investigator: K. Shibata), generous donations
from Sun Microsystems, and NAOJ internal funding. Financial support from the
Spanish Ministerio de Educaci\'on y Ciencia through project
ESP2006-13030-C06-02 is gratefully acknowledged.  \end{acknowledgements}


\end{document}